\documentclass[reprint, superscriptaddress, amsmath, amssymb, aps, prl, floatfix,]{revtex4-2}
\pdfoutput=1

\usepackage{graphicx,
            slashed, 
            subcaption, 
            braket, 
            bm,
            bbm,
            }
\usepackage[compat=1.1.0]{tikz-feynman}
\usepackage[justification=raggedright, singlelinecheck=false]{caption}
\usetikzlibrary{arrows.meta, positioning, decorations.pathmorphing}

\usepackage[pagebackref=false,hidelinks]{hyperref}

\usepackage{soul}
\usepackage[normalem]{ulem}

\newcommand{\beq}{\begin{equation}}
\newcommand{\eeq}{\end{equation}}

\newcommand{\cM}{\mathcal{M}}
\newcommand{\cMbar}{\overline{\mathcal{M}}}
\newcommand{\mmbar}{\cM \text{--} \cMbar}
\newcommand{\cAs}{\mathcal{A}(\overline{\mathcal{M}} \rightarrow \bar{f})}
\newcommand{\cAb}{\mathcal{A}(\mathcal{M} \rightarrow \widetilde f)}

\newcommand{\D}{\mathrm{d}}

\newcommand{\efcut}{\mathcal{E}_\text{f}^\text{min}}

\newcommand{\es}{\mathcal{E}_\text{s}}

\newcommand{\cEs}{\mathcal{E}_\text{s}}
\newcommand{\cEf}{\mathcal{E}_\text{f}}
\newcommand{\escut}{{\cal E}_\text{s}^\text{max}}
\newcommand{\Brs}{\text{BR}_\text{sig}}
\newcommand{\Brb}{\text{BR}_\text{bkg}}

\newcommand{\PC}{P_\text{C}}

\newcommand{\prlsec}[1]{\textit{\textbf{#1---}}}

\begin{document}

\title{Irreducible Bhabha background in the detection of muonium-antimuonium conversion}

\author{Mitrajyoti Ghosh}
\email{mghosh2@fsu.edu}
\affiliation{Department of Physics, Florida State University, 77 Chieftan Way, Tallahassee, FL 32306-4350, USA}

\author{Kevin Liguori}
\email{kliguori@fsu.edu}
\affiliation{Department of Physics, Florida State University, 77 Chieftan Way, Tallahassee, FL 32306-4350, USA}

\author{Takemichi Okui}
\email{tokui@fsu.edu}
\affiliation{Department of Physics, Florida State University, 77 Chieftan Way, Tallahassee, FL 32306-4350, USA}
\affiliation{Theory Center, High Energy Accelerator Research Organization (KEK), 1-1 Oho, Tsukuba, Ibaraki 305-0801, Japan}

\author{Kohsaku Tobioka}
\email{ktobioka@fsu.edu}
\affiliation{Department of Physics, Florida State University, 77 Chieftan Way, Tallahassee, FL 32306-4350, USA}
\affiliation{Theory Center, High Energy Accelerator Research Organization (KEK), 1-1 Oho, Tsukuba, Ibaraki 305-0801, Japan}

\begin{abstract}
Experiments such as MACS and the proposed MACE study muonium-antimuonium conversion by the energies of the final-state $e^\pm$. The $e^+$ and $e^-$ from an antimuonium decay tend to be non-relativistic and relativistic, respectively, and vice versa for muonium. However, these $e^\pm$ can exchange their energies by hard Bhabha scattering, causing muonium to fake an antimuonium decay signal. We compute the rate for this background and find that, while negligible for MACE, it will become larger than the signal for conversion probabilities less than $10^{-18}$. Measuring the helicity of the $e^-$ will reduce this to $10^{-22}$.
\end{abstract}

\maketitle

\prlsec{Introduction}
The conversion of muonium (a bound state of a $\mu^+$ and an $e^-$) to antimuonium ($\mu^-e^+$)~\cite{Pontecorvo:1957cp} is a
clean probe of new physics, being a purely electromagnetic bound state free from hadronic uncertainties.
This idea has long been pursued both theoretically~\cite{Pontecorvo:1957cp, Glashow:1961zz, Feinberg:1961zza, Swartz:1989qz, Hou:1995dg, Horikawa:1995ae, Endo:2020mev, Conlin:2020veq, Han:2021nod, Fukuyama:2021iyw, Petrov:2022wau, Ghosh:2025oju} and experimentally~\cite{Amato:1968xyq, Chang:1989uk, Chatterjee:1992yi, Huber:1988gu, Matthias:1991fw, Willmann:1998gd}. 
The most recent search for such conversion, the MACS experiment~\cite{Willmann:1998gd} at PSI in 1999, observed no events in $\sim 10^{11}$ muonium decays. 
There is also a proposed experiment MACE~\cite{Bai:2024skk} that expects to produce $\sim 10^{14}$ muonia.

Experiments such as MACS and MACE attempt to distinguish the decays of muonium ($\cM = \mu^+e^-$) and antimuonium ($\cMbar = \mu^-e^+$) by the kinematics of the final-state leptons. 
A typical $\cM$ decay produces a fast positron $e^+_{\text{f}}$ (from $\mu^+\to e^+\nu_e\bar\nu_\mu$  with $E_{e^+} \gg m_e$), a slow electron $e^-_{\text{s}}$ (the bound electron, with kinetic energy $\sim m_e\alpha^2 \ll m_e$, with $\alpha = e^2/4\pi$), and two neutrinos.
We denote this final state by $f \equiv e^+_{\text{f}}\, e^-_{\text{s}}\, \bar\nu_\mu \nu_e$. 
A typical $\cMbar$ decay yields instead a slow positron $e^+_{\text{s}}$, a fast electron $e^-_{\text{f}}$, and two neutrinos: $\bar f \equiv e^+_{\text{s}}\, e^-_{\text{f}}\, \nu_\mu \bar\nu_e$. 
An irreducible background arises when the $e^+_\text{f}$ and $e^-_\text{s}$ from an $\cM$ decay undergo hard Bhabha scattering so that $e^+_\text{f}$ and $e^-_\text{s}$ become $e^+_\text{s}$ and $e^-_\text{f}$, mimicking $\bar f$. 
We denote such final state by $\widetilde f \equiv e^+_{\text{s}}\, e^-_{\text{f}}\, \bar\nu_\mu \nu_e$. 
Note that $\widetilde{f}$ is by definition indistinguishable from $\bar f$, provided that the experiment neither distinguishes between $\bar\nu_\mu \nu_e$ and $\nu_\mu \bar\nu_e$ nor measures the helicity of $e_\text{f}^-$ (which is left-handed in $\bar f$). 
Assuming this is the case, as in MACS and MACE, the process $\cM \to \widetilde{f}$ constitutes an \emph{irreducible} background to the signal $\cM \to \cMbar \to \bar{f}$.

In this work, we compute, for the first time, the rate for $\cM \to \widetilde f$. 
We find the relation between the number of muonia and the lowest $\mmbar$ conversion probability that can be probed.
Our results show that, while negligible at the proposed MACE experiment, this Bhabha background will impact the reach of future experiments. 
In particular, for $\mmbar$ conversion probabilities smaller than $10^{-18}$, the Bhabha background is larger than the signal, and hence can no longer be neglected.
We also discuss how measuring the helicity of $e_\text{f}^-$ allows for further discrimination by at least several orders of magnitude. 

Bhabha scattering between $e^+_\text{f}$ and $e^-_\text{s}$ was considered by Feinberg and Weinberg~\cite{Feinberg:1961zza} where it was required that a large energy of more than 10~MeV be transferred from the $e_\text{f}^+$ to $e_\text{s}^-$.
However, they did not impose the final $e^+$ energy to be at the atomic energy scale, $\sim m_e \alpha^2$.
Thus, their background does not mimic $\bar f$ and is reducible.

The Bhabha background considered in this work should not be confused with the \emph{accidental} Bhabha background discussed in~\cite{Willmann:1998gd, Bai:2024skk}, where the $e^+_\text{f}$ from an $\cM$ decay scatters with an $e^-$ in the detector (i.e., not the $e^-_\text{s}$ from the $\cM$ decay) and produces a fast $e^-$ that may be mistaken as coming from an $\cMbar$ decay.
Such accidental Bhabha background is reducible by experimental design. For example, MACE plans to use a pulsed muon beam with late-time windows, tight time-of-flight and energy selection for $e^+_{\text{s}}$, and transverse-momentum requirements on $e^-_{\text{f}}$ to reduce this accidental background. The dominant background at MACE is expected to be ``internal conversion''~\cite{Bai:2024skk} where a $\mu^+$ in the beam decays as $\mu^+ \to e^+_\text{f} e^-_\text{s} e^+ \bar{\nu}_{\mu} \nu_e$ with the extra $e^+$ going undetected. 
We assume this background will be reduced in future experiments, for example, by detecting the extra $e^+$. 

\begin{figure}[t]
    \begin{tikzpicture}[thick, scale=1]

        \begin{feynman}[scale=1]
            \vertex [draw, circle, minimum size=1cm, fill=gray!30] (in) at (4, 0) {\(\overline{\cal M}\)};
            \vertex (p1) at (1, 2);
            \vertex (p2) at (2, 2);
            \vertex (p3) at (2, 0);
            \vertex (o1) at (0,3);
            \vertex (o2) at (0,2);
            \vertex (o3) at (0,1);
            \vertex (o4) at (0,0);
            \vertex (out1) at (-1, 3) {\(\nu_\mu\)};
            \vertex (out2) at (-1, 2) {\(\overline\nu_e\)};
            \vertex (out3) at (-1, 1) {\(e^-_\text{f}\)};
            \vertex (out4) at (-1, 0) {\(e^+_\text{s}\)};

            \diagram* {
                (in) -- [fermion, dashed, quarter right, edge label'=\(\mu^-\)] (p2) -- [dashed] (p1) -- [fermion, dashed, quarter right] (o1) -- [dashed, momentum=\(p_1\)] (out1),
                (out2) --[reversed momentum'=\(p_2\)] (o2) -- [fermion] (p1) -- [fermion, quarter left] (o3) -- [momentum=\(p_\text{f}\)] (out3),
                (in) -- (p3) -- [anti fermion] (o4) -- [momentum=\(p_\text{s}\)] (out4),
            };
        \end{feynman}

        \draw[<-] (4.6, 0) -- (5.1, 0);
        \node at (4.85, 0.25) {$P$};
        
\end{tikzpicture}
    \caption{Diagram for the antimuonium ($\cMbar$) decay part of the signal, not including the conversion ($\cM \to \cMbar$) part. Dashed lines carry muon number, while solid lines electron number. The momenta carried by the outgoing fermion lines are labelled next to the line, and $P$ is the 4-momentum of the $\cMbar$. Crucially,  $e_\text{f}^-$ and $e_\text{s}^+$ are relativistic and non-relativistic, respectively.}
    \label{fig:signal}
\end{figure}

\prlsec{Decay rates of muonium and antimuonium}
The typical kinematics of $\cMbar$ decay, which defines the signal region of phase space, $f$, is shown in Fig.\,\ref{fig:signal}.  Here, the $e^+_\text{s}$ coming from the bound state has kinetic energy $\sim m_e \alpha^2$ while the $e^-_\text{f}$ coming from the $\mu^-$ is relativistic with kinetic energy $\sim m_\mu$. 
To be concrete, we consider the case of pseudo-scalar muonium. 
To calculate the $e$-$\cM$-$\mu$ vertex (the gray blob in Fig.\,\ref{fig:signal}), we follow the treatment of bound states in the appendix of \cite{Ghosh:2025oju}, where, instead of ${\cal O}^\mu = \bar{\mu} \gamma^\mu P_L e$, we use ${\cal O} = \bar{\mu} \gamma_5 e$ to create/annihilate (anti-)muonium.
Working in the muonium rest frame ($P = (m_\mu, \mathbf{0})$ in Fig.\,\ref{fig:signal}), and keeping only the leading term in $m_e/m_\mu$ and/or $\alpha$, we find that the spin-summed squared amplitude is
\begin{equation}\label{eq:amplitude_signal}
    \braket{|{\cAs}|^2} = 2^{13} \pi m_e G_\text{F}^2 a_0^3 \frac{(p_1 \!\cdot p_{\text{f}})(p_2 \!\cdot\! P)}{(1+a_0^2 \bm{p}_{\text{s}}^2)^4} \,,
\end{equation}
where $a_0 \equiv 1/m_e \alpha $ is the Bohr radius of muonium, hence $a_0 |\bm{p}_\text{s}| \sim 1$.
The $m_e$ and $a_0$ dependence in the prefactor can be understood as follows: the spinor for the $e^+_\text{s}$ contributes a factor of $\sqrt{m_e}$, while the momentum space wavefunction of muonium contributes $a_0^{3/2}$ to the amplitude, leading to the $m_e a_0^3$ dependence in the squared amplitude.

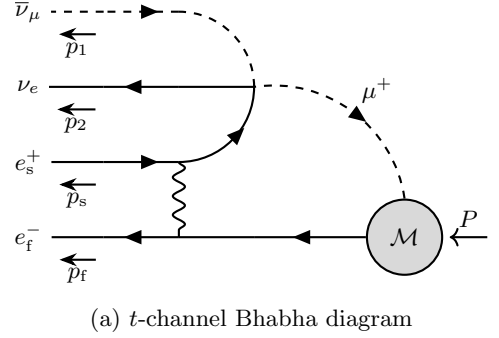
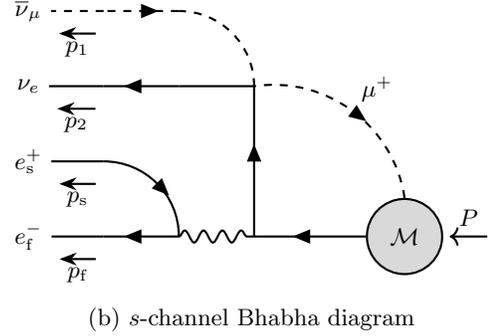
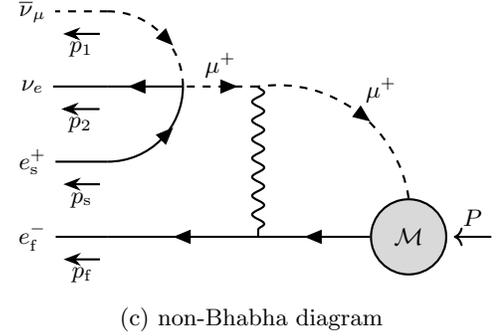
\begin{figure}[t]
\begin{center}    
  \begin{subfigure}[b]{0.48\textwidth}
    \begin{tikzpicture}[thick, scale=1]

        \begin{feynman}[scale=1]
            \vertex [draw, circle, minimum size=1cm, fill=gray!30] (p50) at (5, 0) {\({\cal M}\)};
            \vertex (p00) at (0, 0) {\(e^-_\text{f}\)};
            \vertex (p10) at (1, 0);
            \vertex (p20) at (2, 0);
            \vertex (p30) at (3, 0);
            \vertex (p01) at (0, 1) {\(e^+_\text{s}\)};
            \vertex (p11) at (1, 1);
            \vertex (p21) at (2, 1);
            \vertex (p02) at (0, 2) {\(\nu_e\)};
            \vertex (p12) at (1, 2);
            \vertex (p22) at (2, 2);
            \vertex (p32) at (3, 2);
            \vertex (p03) at (0, 3) {\(\overline\nu_\mu\)};
            \vertex (p13) at (1, 3);
            \vertex (p23) at (2, 3);

            \diagram* {
                (p02) -- [reversed momentum'=\(p_2\)] (p12) -- [anti fermion] (p22) -- (p32) -- [anti fermion, quarter left] (p21) -- [anti fermion] (p11) -- [momentum=\(p_\text{s}\)] (p01),
                
                (p00) -- [reversed momentum'=\(p_\text{f}\)] (p10) -- [anti fermion] (p20) -- (p30) -- [anti fermion] (p50) -- [anti fermion, dashed, quarter right, edge label'=\(\mu^+\)] (p32) -- [dashed, quarter right] (p23) -- [anti fermion, dashed] (p13) -- [dashed, momentum=\(p_1\)] (p03),
            };
            \draw[decorate, decoration={snake, amplitude=0.08cm, segment length=0.25cm}] (p21) -- (p20);
        \end{feynman}

        \draw[<-] (5.6, 0) -- (6.1, 0);
        \node at (5.85, 0.25) {$P$};
        
\end{tikzpicture}
    \caption{\centering $t$-channel Bhabha diagram}
    \label{fig:background_t_bhabha}
  \end{subfigure}
  \vfill
  \begin{subfigure}[b]{0.48\textwidth}
    \begin{tikzpicture}[thick, scale=1]

        \begin{feynman}[scale=1]
            \vertex [draw, circle, minimum size=1cm, fill=gray!30] (p50) at (5, 0) {\({\cal M}\)};
            \vertex (p00) at (0, 0) {\(e^-_\text{f}\)};
            \vertex (p10) at (1, 0);
            \vertex (p20) at (2, 0);
            \vertex (p30) at (3, 0);
            \vertex (p01) at (0, 1) {\(e^+_\text{s}\)};
            \vertex (p11) at (1, 1);
            \vertex (p02) at (0, 2) {\(\nu_e\)};
            \vertex (p12) at (1, 2);
            \vertex (p22) at (2, 2);
            \vertex (p32) at (3, 2);
            \vertex (p03) at (0, 3) {\(\overline\nu_\mu\)};
            \vertex (p13) at (1, 3);
            \vertex (p23) at (2, 3);

            \diagram* {
                (p02) -- [reversed momentum'=\(p_2\)] (p12) -- [anti fermion] (p22) -- (p32) -- [anti fermion] (p30) -- [anti fermion] (p50) -- [anti fermion, dashed, quarter right, edge label'=\(\mu^+\)] (p32) -- [dashed, quarter right] (p23) -- [anti fermion, dashed] (p13) -- [dashed, momentum=\(p_1\)] (p03),
                
                (p00) -- [reversed momentum'=\(p_\text{f}\)] (p10) -- [anti fermion] (p20) -- [anti fermion, quarter right] (p11) -- [momentum=\(p_\text{s}\)] (p01)
            };
            \draw[decorate, decoration={snake, amplitude=0.08cm, segment length=0.25cm}] (p20) -- (p30);
        \end{feynman}

        \draw[<-] (5.6, 0) -- (6.1, 0);
        \node at (5.85, 0.25) {$P$};
        
\end{tikzpicture}
    \caption{\centering $s$-channel Bhabha diagram }
    \label{fig:background_s_bhabha}
  \end{subfigure}
  \vfill
  \begin{subfigure}[b]{0.48\textwidth}
    \begin{tikzpicture}[thick, scale=1]

        \begin{feynman}[scale=1]
            \vertex [draw, circle, minimum size=1cm, fill=gray!30] (in) at (4, 0) {\({\cal M}\)};
            \vertex (p1) at (1, 2);
            \vertex (p2) at (2, 2);
            \vertex (p3) at (2, 0);
            \vertex (o1) at (0,3);
            \vertex (o2) at (0,2);
            \vertex (o3) at (0,1);
            \vertex (o4) at (0,0);
            \vertex (out1) at (-1, 3) {\(\overline\nu_\mu\)};
            \vertex (out2) at (-1, 2) {\(\nu_e\)};
            \vertex (out3) at (-1, 1) {\(e^+_\text{s}\)};
            \vertex (out4) at (-1, 0) {\(e^-_\text{f}\)};

            \diagram* {
                (in) -- [anti fermion, dashed, quarter right, edge label'=\(\mu^+\)] (p2) -- [anti fermion, dashed, edge label'=\(\mu^+\)] (p1) -- [anti fermion, dashed, quarter right] (o1) -- [dashed, momentum=\(p_1\)] (out1),
                (out2) --[reversed momentum'=\(p_2\)] (o2) -- [anti fermion] (p1) -- [anti fermion, quarter left] (o3) -- [momentum=\(p_\text{s}\)] (out3),
                (in) -- [fermion] (p3) -- [fermion] (o4) -- [momentum=\(p_\text{f}\)] (out4),
            };
            \draw[decorate, decoration={snake, amplitude=0.08cm, segment length=0.25cm}] (p2) -- (p3);
        \end{feynman}

        \draw[<-] (4.6, 0) -- (5.1, 0);
        \node at (4.85, 0.25) {$P$};
        
\end{tikzpicture}

% \begin{tikzpicture}[thick, scale=1]

%         \begin{feynman}[scale=1]
%             \vertex [draw, circle, minimum size=1cm, fill=gray!30] (p50) at (5, 0) {\({\cal M}\)};
%             \vertex (p00) at (0, 0) {\(e^-_\text{f}\)};
%             \vertex (p10) at (1, 0);
%             \vertex (p20) at (2, 0);
%             \vertex (p30) at (3, 0);
%             \vertex (p01) at (0, 1) {\(e^+_\text{s}\)};
%             \vertex (p11) at (1, 1);
%             \vertex (p21) at (2, 1);
%             \vertex (p02) at (0, 2) {\(\nu_e\)};
%             \vertex (p12) at (1, 2);
%             \vertex (p22) at (2, 2);
%             \vertex (p32) at (3, 2);
%             \vertex (p03) at (0, 3) {\(\overline\nu_\mu\)};
%             \vertex (p13) at (1, 3);
%             \vertex (p23) at (2, 3);

%             \diagram* {
%                 (p02) -- [reversed momentum'=\(p_2\)] (p12) -- [anti fermion] (p22) -- (p32) -- [anti fermion, quarter left] (p21) -- [anti fermion] (p11) -- [momentum=\(p_\text{s}\)] (p01),
                
%                 (p00) -- [reversed momentum'=\(p_\text{f}\)] (p10) -- [anti fermion] (p20) -- (p30) -- [anti fermion] (p50) -- [anti fermion, dashed, quarter right] (p32) -- [dashed, quarter right] (p23) -- [anti fermion, dashed] (p13) -- [dashed, momentum=\(p_1\)] (p03),
%             };
%             \draw[decorate, decoration={snake, amplitude=0.08cm, segment length=0.25cm}] (p21) -- (4,0);
%         \end{feynman}

%         \draw[<-] (5.6, 0) -- (6.1, 0);
%         \node at (5.85, 0.25) {$P$};
        
% \end{tikzpicture}
    \caption{\centering non-Bhabha diagram}
    \label{fig:background_non_bhabha}
  \end{subfigure}  
\end{center}
\caption{Diagrams for the hard photon exchange background. Note that the final state $e^+_\text{s}$ and $e^-_\text{f}$ have the same four-momenta $p_\text{s}$ and $p_\text{f}$, respectively, as those in the signal (Fig.\,\ref{fig:signal}), and hence this final state is kinematically indistinguishable from that of the signal.}
\label{fig:background}
\end{figure}

Of interest in this work is the case when $\cM$ decays before conversion and undergoes hard scattering that causes the $e^\pm$ to have the ``wrong'' kinematics, that is, $\cM \rightarrow \widetilde f$. 
Fig.\,\ref{fig:background} shows three different 1-loop QED diagrams that contribute to this process.
Figs.\,\ref{fig:background_t_bhabha} and \ref{fig:background_s_bhabha} depict the $t$- and $s$-channel Bhabha diagrams,  respectively. 
The third ``non-Bhabha" diagram, shown in Fig.\,\ref{fig:background_non_bhabha}, is much smaller than the Bhabha diagrams, as we will explain below. Then, ignoring the non-Bhabha diagram, we find that the spin-summed squared amplitude for the irreducible Bhabha background is given by
\begin{equation}\label{eq:amplitude_bhabha}
    \!\!\!
    \braket{|{\cAb}|^2} = 512 \pi G_\text{F}^2 \alpha^3 \frac{(p_1 \!\cdot p_{\text{f}})(p_2 \!\cdot P)}{E_{\text{f}}^2+a_0^2 (\bm{p}_{\text{f}} \cdot \bm{p}_{\text{s}})^2} \,,\!\!
\end{equation}
where $E_\text{f} \gg m_e$ is the energy of the $e^-_\text{f}$.

The amplitude for the non-Bhabha process is suppressed by a factor of $ \sim m_e \alpha / m_\mu$ with respect to the $t$- and $s$-channel Bhabha amplitudes. 
To see why this is the case, note that in both the $t$- and $s$-channel Bhabha diagrams, there is a positron propagator inside the loop which goes as the inverse loop momentum.
However, because the bound-state wave function cuts the integral off at loop momenta $\sim 1/a_0 = m_e \alpha$, the positron propagator contributes a factor of $\sim 1 / m_e \alpha$ to the amplitude. 
In the non-Bhabha diagram, on the other hand, there is a muon propagator outside of the loop instead, which contributes a factor of $\sim 1/m_\mu$ to the amplitude. 
Hence the non-Bhabha amplitude is smaller than the $t$- and $s$-channel Bhabha by $\sim m_e\alpha/m_\mu$.

Using \eqref{eq:amplitude_signal} and \eqref{eq:amplitude_bhabha}, the differential decay rates for the signal and background are 
\begin{equation}\label{eq:rate_signal}
    \begin{aligned}
        &\frac{\D ^3 \Gamma (\overline{\mathcal{M}} \rightarrow \bar{f})}{\D x \, \D \mathcal{E}_{\text{f}} \, \D \mathcal{E}_{\text{s}}} 
        = \frac{m_\mu^5 G_\text{F}^2}{4 \pi^4} \, \mathcal{E}_\text{f}^2 \!\left(1 -\frac{2}{3}\mathcal{E}_{\text{f}}\right)\! \frac{\sqrt{\mathcal{E}_{\text{s}}}}{(1+  \mathcal{E}_{\text{s}})^4} \, 
    \end{aligned}
\end{equation}
and
\begin{equation}\label{eq:rate_bhabha}
    \begin{aligned}
        \frac{\D ^3 \Gamma (\cM \rightarrow \widetilde f)}{\D x \, \D \mathcal{E}_{\text{f}} \, \D \mathcal{E}_{\text{s}}} 
        &= \frac{m_\mu^5 G_\text{F}^2 \alpha^6}{16 \pi^4} \!\left(\frac{m_e}{m_\mu}\right)^{\!\!2}\! \!\left(1 -\frac{2}{3}\mathcal{E}_{\text{f}}\right)\!\frac{\sqrt{\mathcal{E}_{\text{s}}}}{1+  \mathcal{E}_{\text{s}} x^2} \,, 
    \end{aligned}
\end{equation}
respectively, where $\mathcal{E}_{\text{f}} \equiv E_{\text{f}}/(m_\mu / 2)$, $\es \equiv E_{\text{s}}/(m_e \alpha^2 /2 )$, $E_{\text{s}} \sim m_e \alpha^2$ is the kinetic energy of the $e^+_\text{s}$, and 
$x \equiv \cos\theta_\text{fs}$ with $\theta_\text{fs}$ being the angle between $\bm{p}_\text{f}$ and $\bm{p}_\text{s}$. As a sanity check, the total decay rate $\Gamma(\cMbar \rightarrow \bar f)$ is given by
\begin{equation}
    \begin{aligned}
    \Gamma(\cMbar \rightarrow \bar{f}) &= \!\int_{-1}^{1} \!\!\! \D x 
    \!\int_0 ^{\infty} \!\!\!\! \D \mathcal{E}_{\text{s}} 
    \!\int_{0}^{1} \!\! \D \mathcal{E}_{\text{f}} \,\frac{\D ^3 \Gamma (\cMbar \rightarrow \bar{f})}{\D x \, \D {\cal E}_{\text{f}} \, \D {\cal E}_{\text{s}}} \\ 
    &= \frac{G_\text{F}^2 m_\mu^5}{192 \pi^3} \equiv \Gamma_0\,,
    \end{aligned}
\end{equation}
which is the total decay rate of the muon, as expected. We have extended the upper bound of the  ${\cal E}_\text{s}$ integral to infinity because the integrand dies off quickly for $\cEs \gtrsim 1$. 
Similarly, the lower bound on the $\cEf$ integral has been extended from $2 m_e / m_\mu$ to $0$.
These extensions are consistent with our approximation of working at leading order in $m_e / m_\mu$.

\prlsec{Results}
Using~\eqref{eq:rate_signal} and~\eqref{eq:rate_bhabha}, 
the differential branching fractions $\Brs$ and $\Brb$ for the signal ($\cM \to \cMbar \to \bar{f}$) and irreducible Bhabha background ($\cM \to \widetilde{f}$), respectively, are given by
\begin{equation}\label{eq:dbrdes_sig_full}
    \begin{aligned}
        \frac{\D^3 \Brs}{\D x \, \D \cEf \, \D \cEs} &\equiv \PC \frac{\D^3 \Gamma(\cMbar \rightarrow \bar{f})}{\Gamma_0 \, \D x \, \D \cEf \, \D \cEs} \,,
    \end{aligned}
\end{equation}
and
\begin{equation}\label{eq:dbrdes_bkg_full}
    \begin{aligned}
        \frac{\D^3 \Brb}{\D x \, \D \cEf \, \D \cEs} &\equiv \left(1 - \PC\right)  \frac{\D^3
        \Gamma(\cM \rightarrow \widetilde f)}{\Gamma_0 \, \D x \, \D \cEf \, \D \cEs} \\ 
        &\simeq \frac{\D^3 \Gamma(\cM \to \widetilde f)}{\Gamma_0 \, \D x \, \D \cEf \, \D \cEs} \,,
    \end{aligned}
\end{equation}
where $\PC \equiv P(\cM \rightarrow \cMbar) \ll 1$ is the $\mmbar$ conversion probability. 

Experimentally, to ensure the $e^+_\text{s}$ and $e^-_\text{f}$ are sufficiently slow and fast, respectively, cuts are imposed on $\cEs$ and $\cEf$. Specifically, we require $0 \leq \cEs \leq \escut$ and $\efcut \leq \cEf \leq 1$.
Integrating \eqref{eq:dbrdes_sig_full} and \eqref{eq:dbrdes_bkg_full} over $\cEs$ and $\cEf$ in these ranges and $x \in [-1, 1]$, 
we find the branching fractions are:
\begin{equation}\label{eq:brs}
\Brs = \frac{2}{\pi} \PC \, \overline{\mathcal{F}}(\efcut) \, \overline{\mathcal{S}}(\escut)  
\end{equation}
and
\begin{equation}\label{eq:brb}
\Brb 
= \frac{8 \alpha^6}{\pi} \!\left(\frac{m_e}{m_\mu}\right)^{\!\!2} \, \widetilde{\cal F}(\efcut) \, \widetilde{\mathcal{S}}(\escut) \,,
\end{equation}
where
\begin{equation}\label{eq:sfbar}
    \begin{aligned}
        \overline{\mathcal{F}}(z) &\equiv (1-z)(1+z+z^2-z^3) \,,
        \\
        \overline{\mathcal{S}}(z) &\equiv \arctan\sqrt{z} - \dfrac{\sqrt{z}(1+\frac{1}{3}z)(1-3z)}{(1+z)^3} \,,
    \end{aligned}
\end{equation}
and
\begin{equation}\label{eq:sftilde}
    \begin{aligned}
        \widetilde{\mathcal{F}}(z) &\equiv (1 - z)(2 - z) \,,
        \\
        \widetilde{\cal S}(z) &\equiv (1 + z)\arctan\sqrt{z} -\sqrt{z} \,.
    \end{aligned}
\end{equation}
Setting the lower limit of $\cEs$ to 0 as above is a good approximation experimentally (e.g., see \cite{Bai:2024skk, Lu:2025col}), and
we assume that this will still be the case for future experiments. 
Similarly, we have integrated over all $\theta_\text{fs}$ (i.e. all $x$), assuming that it is not measured, as in MACS and MACE\@.  
Since the $\theta_\text{fs}$ dependence of the background, as seen in Eq.\,\eqref{eq:rate_bhabha}, is very mild, imposing cuts on $\theta_\text{fs}$ will not change our conclusion.

Now, using \eqref{eq:brs} and \eqref{eq:brb}, the ratio of the signal to the irreducible Bhabha background is given by
\begin{equation}\label{eq:R}
    \begin{aligned}
        {\cal R} &\equiv \frac{\,\,\Brs\,\,}{\,\, \Brb \,\,} = \frac{\PC}{4 \alpha^6} \!\left( \frac{m_\mu}{m_e}\right)^{\!\!2} \frac{\overline{{\cal F}} (\efcut)}{\widetilde{\cal F}(\efcut)}  \frac{\overline{{\cal S}} (\escut)}{\widetilde {\cal S}(\escut)}\, \\
        &= \frac{\PC}{{1.4 \times 10^{-17}}} \,  \frac{\overline{{\cal F}} (\efcut)}{\widetilde{\cal F}(\efcut)}  \frac{\overline{{\cal S}} (\escut)}{\widetilde {\cal S}(\escut)}\, .
    \end{aligned}
\end{equation}
Because $\overline{\mathcal{F}} \, \overline{\mathcal{S}} / \widetilde{\mathcal{F}} \, \widetilde{\mathcal{S}}$ has a maximum value of 16, $\mathcal{R} \leq (1.1 \times 10^{18}) \, \PC$. Thus, for conversion probabilities smaller than $0.88 \times 10^{-18}$ the branching fraction for the Bhabha background becomes greater than that of the signal, regardless of the cuts $\efcut$ and $\escut$. 

\begin{figure}[t]
    \begin{center}
    \includegraphics[width=0.45\textwidth]{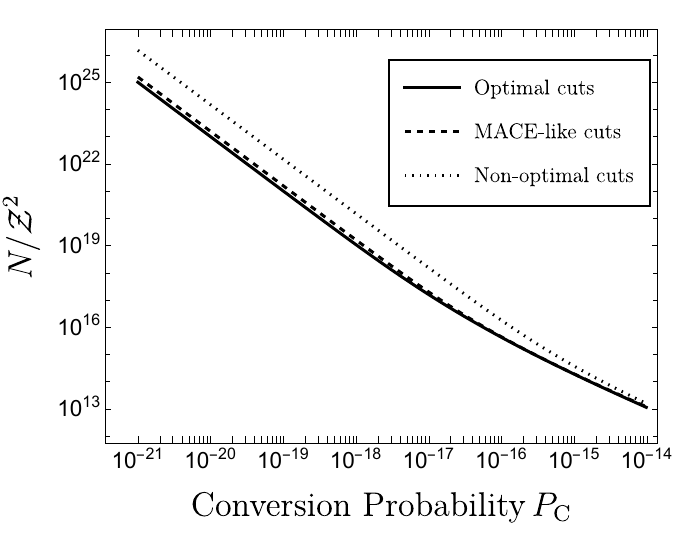}
    \end{center}
    \caption{
    $N / \mathcal{Z}^2$ as a function of $\PC$ for different scenarios. The solid line uses the optimal cuts as shown in Fig.~\ref{fig:esfminmax}, i.e., $N=N_\text{min}$. The dashed line uses the fixed cuts similar to those proposed by MACE~\cite{Bai:2024skk} ($\efcut = 0.4$ and $\escut = 1.5$). The dotted line uses poorly chosen fixed cuts $\efcut = 0.1$ and $\escut = 9$ to illustrate that non-optimal choices do lead to a higher $N$, but not significantly higher.}
    \label{fig:noverz2}
\end{figure}

\begin{figure}[t]
    \begin{center}            \includegraphics[width=0.48\textwidth]{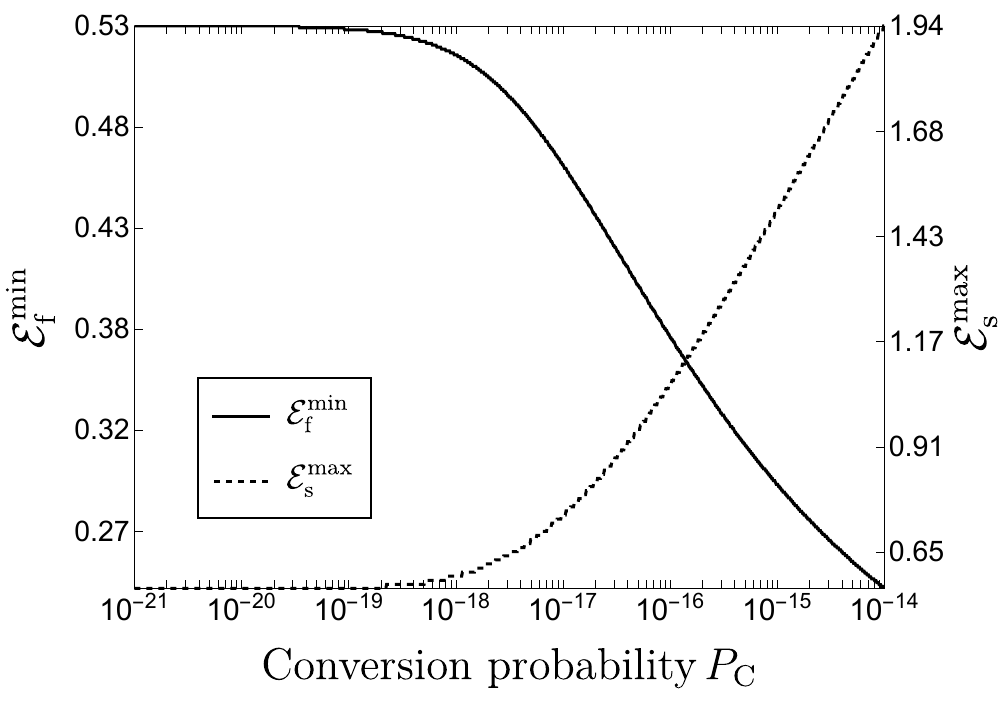}
    \end{center}
\caption{
The optimal values of $\efcut$ (solid) and $\escut$ (dashed) that yield $N = N_\text{min}$ for a given $\PC$.}
    \label{fig:esfminmax}
\end{figure}

Armed with an expression for the branching fraction of the irreducible Bhabha background, we now investigate how many muonia are needed in an experiment to probe a desired $\PC$. 
Let us consider an $\mmbar$ conversion experiment with $N$ muonia. The expected number of signal events is $N \, \Brs$, while the expected number of irreducible Bhabha background events is $N \, \Brb$. 
We adopt the expected discovery significance ${\cal Z}$ given by~\cite{Cowan:2010js}
\begin{equation}\label{eq:significance}
    {\cal Z} = \sqrt{2 N \, \Brb \big[(1+{\cal R})\ln\left(1+{\cal R}\right)- {\cal R}  \big]}\,.
\end{equation}
We can use \eqref{eq:significance} to find the number of muonia $N$ as a function of $\efcut$, $\escut$, and $\PC$ for a given $\mathcal{Z}$. 
Minimizing $N$ with respect to the cuts yields the minimum number $N_\text{min}$ of muonia necessary to be sensitive to a conversion probability of $\PC$ at significance $\mathcal{Z}$. 
This is plotted as a solid line in Fig.\,\ref{fig:noverz2} which shows $N/\mathcal{Z}^2$ vs $\PC$.
The optimal cuts (as functions of $\PC$) that yield $N_\text{min}$ are plotted in Fig.~\ref{fig:esfminmax}. 
While these cuts do minimize the number of muonia required, it is worthwhile to note that, for reasonable cuts similar to those used in MACS and the proposed MACE (plotted as a dashed line in Fig.\,\ref{fig:noverz2}), the dependence of $N$ on these cuts is very mild. Even a poor choice of cuts (the dotted line in Fig.\,\ref{fig:noverz2}) does not significantly diminish the sensitivity.
The takeaway here is that one cannot use kinematical cuts to reduce the Bhabha background.

%\prlsec{MAGA: Make Antimuonium Great Again by Helicity}
\prlsec{Further discrimination by helicity}
Because $e_\text{f}^- \in \bar f$ is left-handed, this offers the possibility of using the helicity of $e_\text{f}^- \in \widetilde f$ to reduce the Bhabha background. 
This reduction will be significant, because, 
even though we summed over the $e^-_\text{f}$ helicities in~\eqref{eq:amplitude_bhabha}, the entire contribution is from a right-handed $e^-_\text{f}$ at this order in $m_e / m_\mu$ and $\alpha$.
The left-handed $e^-_\text{f}$ contribution completely cancels between the $t$- and $s$-channel diagrams.
Therefore, if the experiment can measure the $e^-_\text{f}$ helicity, we expect that the signal-to-background ratio will be enhanced by a factor of at least $\sim (m_\mu / m_e)^2 \sim 10^4$, i.e., the signal will remain larger than the background for $\PC \gtrsim 10^{-22}$ instead of $10^{-18}$.

\prlsec{Conclusion}
We have computed the rate for the irreducible Bhabha background to muonium-antimuonium conversion.
While Bhabha scattering can give rise to both reducible and irreducible backgrounds, our results focus on the irreducible part of Bhabha scattering where the final state $e^\pm$ are in the same region of the phase space as those of the signal. We obtained the minimum number of muonia necessary to probe a given $\PC$, and found this to be only mildly sensitive to the cuts $\efcut$ and $\escut$ on the energies of the $e^\pm$. Furthermore, we find that for $\PC \leq 10^{-18}$, the Bhabha background is larger than the signal, irrespective of the cuts $\efcut$ and $\escut$, confirming the irreducibility of this background in the absence of neutrino and helicity detection. This also motivates us to explore new detection schemes that do not suffer from this background. For example, perhaps direct detection of the conversion is possible by ionizing $\cMbar$ into $\mu^-e^+$ to distinguish from $\mu^+$ from $\cM$ or the $\mu^+$ beam.

\prlsec{Acknowledgments}
This work was supported in part by the US Department of Energy grant DE-SC0010102, and is supported in part by the FSU Bridge Funding 047302.

\bibliography{mybib}
\bibliographystyle{JHEP}

\end{document}